\documentclass{bioinfo}
\copyrightyear{2019}
\pubyear{2019}

\usepackage{natbib}
\begin{document}
\firstpage{1}

\title{Identifying centromeric satellites with dna-brnn}
\author[Li]{Heng Li$^{1,2,3}$}
\address{$^1$ Department of data sciences, Dana-Farber Cancer Institute, 450 Brookline Ave, Boston, MA 02115, USA\\
$^2$ Department of biomedical informatics, Harvard Medical School, 25 Shattuck Street, Boston, MA 02115, USA\\
$^3$ Broad Institute, 415 Main St, Cambridge, MA 02142, USA}

\maketitle

\begin{abstract}

\section{Summary:} Human alpha satellite and satellite 2/3 contribute to
several percent of the human genome. However, identifying these sequences with
traditional algorithms is computationally intensive. Here we develop dna-brnn,
a recurrent neural network to learn the sequences of the two classes of
centromeric repeats. It achieves high similarity to RepeatMasker and is times
faster. Dna-brnn explores a novel application of deep learning and may
accelerate the study of the evolution of the two repeat classes.

\section{Availability and implementation:}
\href{https://github.com/lh3/dna-nn}{https://github.com/lh3/dna-nn}

\section{Contact:} hli@jimmy.harvard.edu
\end{abstract}

\section{Introduction}

Eukaryotic centromeres consist of huge arrays of tandem repeats, termed
\emph{satellite DNA}~\citep{Garrido-Ramos:2017aa}. In human, the two largest
classes of centromeric satellites are alpha satellite (alphoid) with a 171bp
repeat unit, and satellite II/III (hsat2,3) composed of diverse variations of
the ${\tt ATTCC}$ motif. They are totaled a couple of hundred megabases in
length~\citep{Schneider:2017aa}. Both alphoid and hsat2,3 can be identified
with RepeatMasker~\citep{Tarailo-Graovac:2009aa}, which is alignment based and
uses the TRF tandem repeat finder~\citep{Benson:1999aa}. However, RepeatMasker
is inefficient. Annotating a human long-read assembly may take days; annotating
high-coverage sequence reads is practically infeasible. In addition,
RepeatMasker requires RepBase~\citep{Kapitonov:2008aa}, which is not
commercially free. This further limits its uses.

We reduce repeat annotation to a classification problem and solve the problem
with a recurrent neural network (RNN), which can be thought as an extension to
non-profile hidden Markov model but with long-range memory. Because the repeat
units of alphoid and hsat2,3 are short, RNN can ``memorize'' their sequences with a
small network and achieve high performance.

\begin{methods}
\section{Methods}

\subsection{The dna-brnn model}

Given $m$ types of non-overlapping features on a DNA sequence, we can label
each base with number $0,\ldots,m$, where `0' stands for a null-feature.
\mbox{Dna-brnn} learns how to label a DNA sequence. Its overall architecture
(Fig.~\ref{fig:model}) is similar to an ordinary bidirected RNN
(BRNN), except that \mbox{dna-brnn} feeds the reverse complement sequence to the
opposite array of Gated Recurrent Units (GRUs) and that it ties the weights in both directions. Dna-brnn is
strand symmetric in that the network output is the same regardless of the input
DNA strand. The strand symmetry helps accuracy~\citep{Shrikumar103663}.
%is not just a nice theoretical property.
Without weight sharing between the two strands, we will end up with a model
with twice as many parameters but 16\% worse training cost (averaged in 10
runs).

In theory, we can directly apply dna-brnn to arbitrarily long sequences.
However, given a single sequence or multiple sequences of variable lengths, it
is non-trivial to implement advanced parallelization techniques and without
parallelization, the practical performance would be tens of times slower. As a
tradeoff, we apply dna-brnn to 150bp subsequences and discard information in a
longer range.

To identify satellites, we assign label `1' to hsat2,3 and label `2' to
alphoid. The size of the GRU hidden vector is 32. There are $<$5000 free
parameters in such a model.

\begin{figure}[tb]
\centering
\includegraphics[width=.45\textwidth]{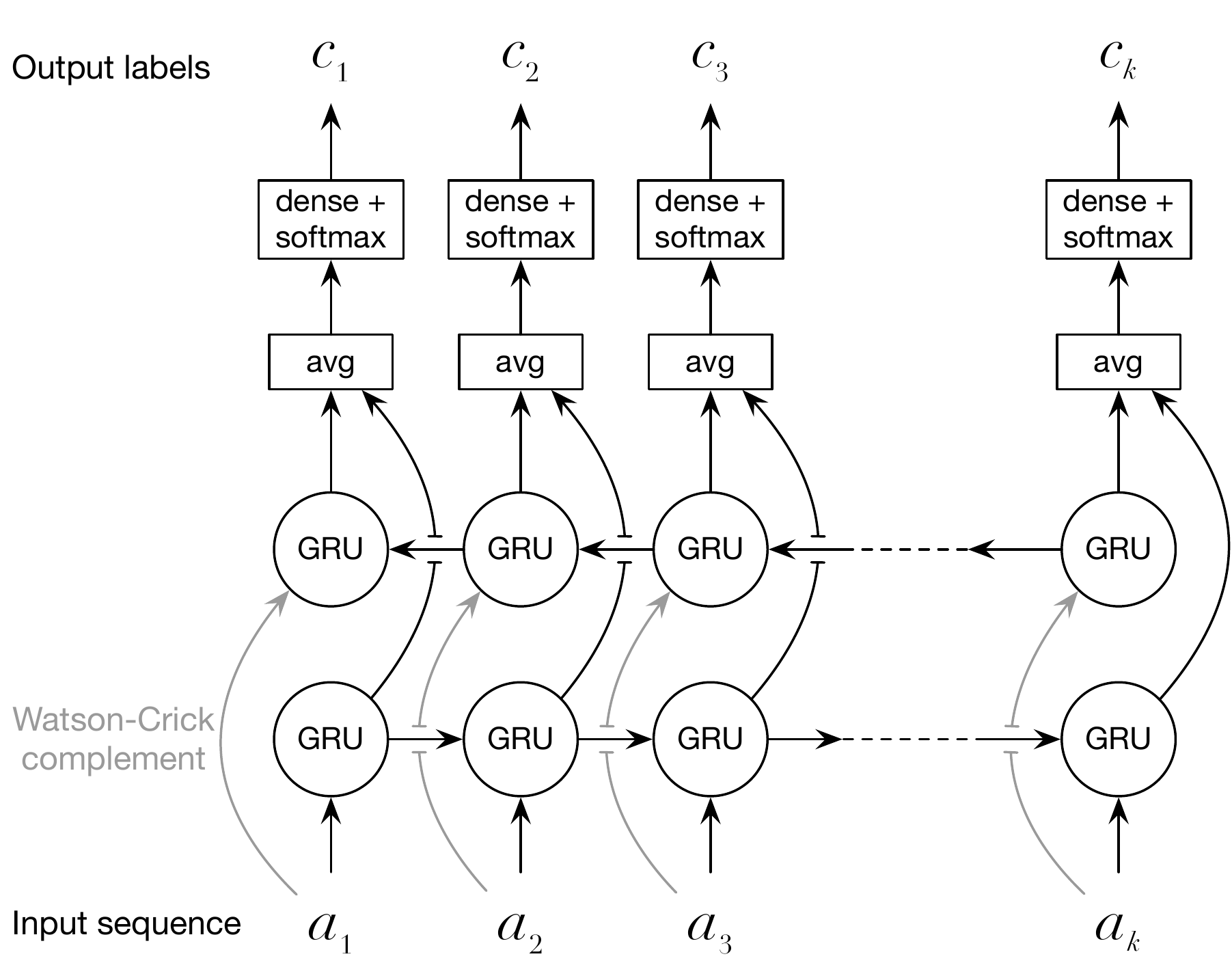}
\caption{The dna-brnn model. Dna-brnn takes a $k$-long one-hot encoded DNA
sequence as input. It feeds the input and its reverse complement to two GRU
arrays running in the opposite directions. At each
position, dna-brnn averages the two GRU output vectors, transforms the
dimension of the average with a dense layer and applies softmax. The final
output is the predicted distribution of labels for each input base. All GRUs in
both directions share the same weights.}\label{fig:model}
\end{figure}

\subsection{Training and prediction}

In training, we randomly sampled 256 subsequences of 150bp in length and updated
the model weights with RMSprop on this minibatch. To reduce overfitting, we
randomly dropped out 25\% elements in the hidden vectors. We terminated
training after processing 250Mb randomly sampled bases. We generated 10 models
with different random seeds and manually selected the one with the best accuracy on the
validation data.

On prediction, we run the model in each 150bp long sliding window with 50bp
overlap. In each window, the label with the highest probability is taken as the
preliminary prediction. In an overlap between two adjacent windows, the label
with higher probability is taken as the prediction. Such a prediction
algorithm works well in long arrays of satellites. However, it occasionally
identifies satellites of a few bases when there is competing evidence. To
address this issue, we propose a post-processing step.

With the previous algorithm, we can predict label $c_i$ and its
probability $p_i$ at each sequence position $i$. We
introduce a score $s_i$ which is computed as
\begin{equation}\label{eq:sc}
t_i=\log\frac{\min(p_i,0.99)}{1-\min(p_i,0.99)},\,\,\,
s_i=\left\{\begin{array}{ll}
t_i & (c_i>0) \\
-10t_i & (c_i=0)\\
\end{array}\right.
\end{equation}
Here $s_i$ is usually positive at a predicted satellite base and negative at a
non-satellite base. Let $S_{a,b}=\sum_{i=a}^b s_i$
be the sum of scores over segment $[a,b]$. Intuitively, we say $[a,b]$ is
maximal if it cannot be lengthened or shortened without reducing $S_{a,b}$.
\citet{DBLP:conf/ismb/RuzzoT99} gave a rigorous definition of maximal scoring
segment (MSS) and a linear algorithm to find all of them. By default,
dna-brnn takes an MSS longer than 50bp as a satellite segment. The use of MSS
effectively clusters fragmented satellite predictions and improves the accuracy
in practice.

\subsection{Training and testing data}

The training data come from three sources: chromosome 11, annotated alphoids in
the reference genome and the decoy sequences, all for GRCh37.  RepeatMasker
annotations on GRCh37 were acquired from the UCSC Genome Browser.  Repeats on
the GRCh37 decoy were obtained by us with RepeatMasker (v4.0.8 with rmblast-2.6.0+ and the human section of
RepBase-23.11). RepeatMasker may annotate hsat2,3 as
`HSATII', `(ATTCC)n', `(GGAAT)n', `(ATTCCATTCC)n' or other rotations of the
ATTCC motif. We combined all such repeats into hsat2,3. We take the
RepeatMasker labeling as the ground truth.

For validation, we annotated the GRCh38 decoy sequences~\citep{Mallick:2016aa}
with RepeatMasker and used that to tune hyperparameters such as the size of
GRU and non-model parameters in Eq.~(\ref{eq:sc}), and to evaluate the effect
of random initialization. For testing, we annotated the CHM1 assembly
(AC:GCA\_001297185.1) with RepeatMasker as well. Testing data do not overlap
training or validation data.

For measuring the speed of RepeatMasker, we used a much smaller repeat
database, composed of seven sequences (`HSATII', `ALR', `ALR\_', `ALRa',
`ALRa\_', `ALRb' and `ALRb\_') extracted from the prepared RepeatMasker
database. We used option `-frag 300000 -no\_is' as we found this achieves the
best performance. The result obtained with a smaller database is slightly
different from that with a full database because RepeatMasker resolves
overlapping hits differently.

\subsection{Implementation}

Unlike mainstream deep learning tools which are written in Python and depend on
heavy frameworks such as TensorFlow, dna-brnn is implemented in C, on top of
the lightweight KANN framework that we developed. KANN implements generic
computation graphs. It uses CPU only, supports parallelization and has no
external dependencies. This makes dna-brnn easily deployed without requiring
special hardware or software settings.

%We have also implemented dna-cnn, a strand-symmetric convolutional neural network
%for a similar task. It predicts the fraction of each repeat class on a
%fixed-length sequence. Dna-cnn is faster than dna-brnn and can also achieve
%high similarity to RepeatMasker. We focus on dna-brnn here because it outputs
%per-base annotation.

\end{methods}

\section{Results}

Training dna-brnn takes 6.7 wall-clock minutes using 16 CPUs; predicting labels
for the full CHM1 assembly takes 56 minutes. With 16 CPUs, RepeatMasker is 5.3 times
as slow in CPU time, but 17 times as slow in real time, likely because it
invokes large disk I/O and runs on a single CPU to collate results. Table~1 shows the
testing accuracy with different prediction strategies. Applying MSS clustering
improves both FNR and FPR. We use the `mss:Y, minLen:50' setting in the rest of
this section.

%On the testing data, dnr-brnn predicts 70kb hsat2,3 satellites not annotated
%by RepeatMasker. We looked at them by eye and had the impression that most of
%them are divergent copies of (ATTCC)n. Some of them are annotated as other
%types of tandem repeats; some are missed. The hsat2,3 FPR of dna-brnn should be
%lower considering that the RepeatMasker annotation may be incomplete.

\begin{table}[tb]
\processtable{Evaluation of dna-brnn accuracy}
{\footnotesize
\begin{tabular}{p{3.3cm}rlrl}
\toprule\\[-1.5em]
& \multicolumn{2}{c}{\vspace{0.3em}alphoid} & \multicolumn{2}{c}{hsat2,3} \\
\cline{2-5}\\[-0.7em]
Setting & \multicolumn{1}{c}{FNR} & \multicolumn{1}{c}{FPR} & \multicolumn{1}{c}{FNR} & \multicolumn{1}{c}{FPR} \\[-0.5em]
\midrule
mss:N, minLen:0 & 0.42\% & 1 / 9952  & 0.42\% & 1 / 4086 \\
mss:N, minLen:50& 0.59\% & 1 / 28908 & 0.68\% & 1 / 4639 \\
mss:Y, minLen:50& 0.33\% & 1 / 44095 & 0.30\% & 1 / 4370 \\
mss:Y, minLen:200&0.36\% & 1 / 60078 & 0.50\% & 1 / 6010 \\
mss:Y, minLen:500&0.48\% & 1 / 69825 & 0.85\% & 1 / 10505 \\
\botrule
\end{tabular}
}{RepeatMasker annotations on the CHM1 assembly (3.0Gb in total, including 55Mb alphoid and 50Mb hsat2,3)
are taken as the ground truth. `mss': whether to cluster
predictions with maximal scoring segments.  `minLen': minimum satellite length.
`FNR': false negative rate, the fraction of RepeatMasker annotated bases being missed
by dna-brnn. `FPR': false positive rate, the fraction of non-satellite bases
being predicted as satellite by dna-brnn. A format `$1/x$' in the table implies
one false positive prediction per $x$-bp.}\label{tab:eval}
\end{table}

Dna-brnn takes $\sim$1.5 days on 16 threads to process whole-genome short or long
reads sequenced to 30-fold coverage. For the NA24385 CCS data
set~\citep{Wenger519025}, 2.91\% of bases are alphoid and 2.56\% are hsat2,3.
If we assume the human genome is 3Gb in size, these two classes of satellites
amount to 164Mb per haploid genome. The CHM1 assembly
contains 105Mb hsat2,3 and alphoid, though 70\% of them are in short
contigs isolated from non-repetitive regions. In the reference
genome GRCh37, both classes are significantly depleted ($<$0.3\% of the
genome).  GRCh38 includes computationally generated alphoids but still lacks
hsat2,3 ($<$0.1\%). Partly due to this, 82\% of human novel sequences found
by~\citet{Sherman:2018aa} are hsat2,3.  There are significantly less novel
sequences in euchromatin.

We have also trained dna-brnn to identify the Alu repeats to high accuracy.
Learning beta satellites, another class of centromeric repeats, is harder. We
can only achieve moderate accuracy with larger hidden layers. Dna-brnn
fails to learn the L1 repeats, which are longer, more divergent and
more fragmented. We are not sure if this is caused by the limited capacity of
dna-brnn or by innate ambiguity in the RepeatMasker annotation.

\section{Conclusion}

Dna-brnn is a fast and handy tool to annotate centromeric satellites on
high-throughput sequence data and may help biologists to understand the
evolution of these repeats. Dna-brnn is also a general approach to modeling DNA
sequences. It can potentially learn other sequence features and can be easily
adapted to different types of sequence classification problems.

\section*{Acknowledgement}

We thank the second anonymous reviewer for pointing out an issue with our
running RepeatMasker, which led to unfair performance comparison in an earlier
version of this manuscript.

\paragraph{Funding\textcolon} NHGRI R01-HG010040

\bibliography{dna-nn}

\end{document}